\documentclass[12pt]{article}
\def\be{\begin{equation}}
\def\ee{\end{equation}}
\def\la{\label}
\def\bea{\begin{eqnarray}}
\def\eea{\end{eqnarray}}
\def\non{\nonumber}
\def\ci{\cite}
\def\la{\label}
\def\bib{\bibitem}

\def\Lm{\Lambda}
\def\le{\left}
\def\ri{\right}

\def\ga{\gamma}

\def\Omp{\Omega_\phi}

\def\rp{\rho_\phi}
\def\rm{\rho_m}
\def\wp{w_\phi}

\def\s8{\sigma_8}

\def\fr{\frac}

\usepackage{graphicx}

\begin{document}

\begin{center}
   {\Large \bf  Cosmology for Scalar Fields with  Negative
    Potentials and $\wp<-1$ }

\end{center}

\vspace*{0.5cm}

\begin{center}
{\bf A. de la Macorra$^a$\footnote{e-mail: macorra@fisica.unam.mx} and G.
 Germ\'an$^b$\footnote{e-mail: gabriel@fis.unam.mx}
}
\end{center}


\begin{center}
{\small
\begin{tabular}{c}
$^a$ Instituto de F\'{\i}sica, UNAM\\
Apdo. Postal 20-364, 01000  M\'exico D.F., M\'exico\\
$^b$ Centro de Ciencias F\'{\i}sicas,
Universidad Nacional Aut{\'o}noma de M{\'e}xico,\\
Apartado Postal 48-3, 62251 Cuernavaca, Morelos,  M{\'e}xico\\
\\
\end{tabular}
}
\end{center}


\begin{center}
{\bf ABSTRACT}
\end{center}
\small{  We study the cosmology of canonically normalized scalar
fields  that lead to an equation of state parameter
of $w_\phi=p_\phi/\rho_\phi<-1$  without
violating  the weak energy condition: $\rho=\Sigma_i\rho_i \geq 0$
and $\rho_i+p_i\geq 0$. This kind of behavior
requires a negative scalar potential $V$, widely predicted in
particle physics.  We show that the energy
density $\rho_\phi=E_k+V$ takes negative values with an
equation of state with $w_\phi < -1$. However,
the net effect of the $\phi$ field on the scale factor
is to decelerate it giving a total equation
of state parameter $w=p/\rho> w_b=p_b/\rho_b$, where
$\rho_b$ stands for any kind of energy density
with $-1\leq w_b \leq 1$, such as radiation, matter,
cosmological constant or other scalar field
with a potential $V\geq0$.

The fact that $\rho_\phi<0$ allows, at least in principle,
to have a small cosmological constant or quintessence today
as the cancellation of high energy scales such as the electroweak
or susy breaking scale. While $V$ is negative
$|\rho_\phi| $ is smaller than the sum of all other energy densities
regardless of the functional form of the potential $V$.

We show that the existence of a   negative
potential leads, inevitable, to a  collapsing universe, i.e.
to a would be "big crunch". In this picture we would still be living
in the expanding universe.

}

\vspace*{.1 cm}

\noindent \rule[.1in]{14.5cm}{.002in}

\thispagestyle{empty}

\setcounter{page}{0} \vfill\eject

\section{Introduction}

The recent cosmological observations show that the universe is in
an accelerating epoch.  These observations, together with CMB
data, require a dark energy density of $\Omega_{de} \simeq 0.7$
and negative equation of state $w_{de}=p_{de}/\rho_{de}<0$
\ci{CMBR,SN1a}. The data suggest that $w_{de}<-2/3$ and the best
fit models have $w_{de}$ very close to minus one   including the
regions around $w_{de}=-1$. Even more, the existing data do not
ruled out a $w_{de}<-1$ \ci{SN1a}. Causality arguments seem to
require $w>-1$ but this is only the case for constant $w$ and in
fact it is possible to have $|w|>1$ without violating causality
\ci{ax.vuc}. The possibility of having $w<-1$ was raised by
\ci{cald}-\ci{mcin} where they studied a phantom energy density
with $\rho_{ph}>0$ and $w_{ph}<-1$. From a particle point of view
it is unclear how such field can be obtained and it was suggested
to have a scalar field with a negative kinetic energy \ci{cald},
\ci{phantom}. In this work, we will concentrate on scalar fields
with canonical kinetic terms ($E_k\geq0$) that lead dynamically to
regions with $|w|>1$.

The cosmology of negative potential has become increasingly
interesting since many particle physics models and string theory
predict anti-de-Sitter "ADS" spaces. The cyclic universe \ci{cyc}
 and the ekpyrotic \ci{ekp}
universe models use a negative potential and describe  a universe
that goes from a contracting phase (leading to a big crunch) to an
accelerating universe. The main idea is to give an alternative
theory to the big bang singularity. However, it is not yet clear
the process of the transition between contracting and accelerating
phases.

Scalar fields are widely obtained in particle physics, and they
can be either fundamental fields or compose fields (fermion
condensates) \ci{chris1,ax.mod}. They can be  used to describe the
cosmological constant as quintessence \ci{tracker} and in all
these models the scalar fields have a positive semi-definite
potential $V$ \ci{generic, mio.scalar}. However, negative
potential are widely predicted by string theory or  by
supergravity (sugra) models. The sugra potential is
\be
V=e^K\le[F_i(K^{-1})^i_jF^j-3|W|^2\ri]
\ee
with $K(\phi,\bar\phi)$ the Kahler potential, $W(\phi)$ the superpotential
and $F_i=K_iW+W_i,\;W_i\equiv dW/d\phi_i$. It is clear that the general
vacua of sugra  has a negative potential and a lot of model building (in many
cases not quite self consistent) has to be done to get a potential
$V\geq 0$.

Negative potentials were studied in \ci{linde} but in the absence
of a background density and in a model dependent analysis. Here, we will study the
cosmology of negative potentials in the presence
of a background energy density, we do a model independent analysis of the
evolution of the scalar field and we give some specific examples.
We will show that these models lead
to an equation of state with $|\wp|>1$ in a natural dynamical way.
The stage where  $\wp < -1$  has a negative
energy density $\rho_\phi <0$ but a positive $\rp+p_\phi=2E_k\geq
0$. These models  lead inevitable to a collapsing universe and to
a would be "big crunch" contrary to phantom fields. Indeed,
phantom physics suggest that the final stage of the universe has a
divergence of the scale factor  at finite time ($a (t) \rightarrow
\infty)$ \ci{mcin} called a "big smash" while
 a "big crunch" ends in a singularity $a(t) \rightarrow 0$, $\rho\rightarrow \infty$.

We do not worry about a big
crunch as we do not worry about a  big bang since we are working
with an effective theory and we would expect that a fundamental
theory like strings could have a graceful exit to the problem of
singularities. The behavior of the universe approaching
the  big crunch singularity is just as bad  as the big bang
singularity since the universe has, in general, a time inversion symmetry $t_0+t\rightarrow
t_0-t$ around the point where the Hubble constant is zero  $H(t_0)=0$. So,
whatever the solution to the big bang singularity will also solve the big
crunch singularity. The evolution of the universe outside the
possible singularities  may well be described by the effective
field theory and general relativity.

An accelerating universe requires a dominating energy density with
$w_i<-1/3$, (or more exact $\Omega_i w_i<-1/3,\;\Omega_i=\rho_i/\rho_{Tot}$),
and therefore we could naively think that a $w_i<-1$ will increase
the acceleration of the
universe. However, a canonical normalized scalar field
with $\wp<-1$ decelerates the universe
because in this region the energy density is negative, as we will show later,
and its contribution to the acceleration
of the scale factor  given by
\be
\ddot{a}\sim - (\rho_\phi+3p_\phi)=-\rho_\phi(1+3\wp)
\ee
is negative, thus decelerating the expansion of the universe.

In particle physics the prediction of many scalar fields is
quite common. It is certainly more natural to assume
the existence of more than one scalar field which
will, in general, have different potentials
and different cosmological evolutions. So, can
quintessence  be the sum of two or more different kinds of scalar fields?
The answer is clearly yes. There is no theoretical or observational
argument against it.

The interesting point is that negative potentials allow, al least
in principle, to explain the cosmological scale in terms of a high
energy scale (e.g. electroweak  or susy breaking scale) since the
contribution of a negative potential will cancel that of a
positive potential rendering a finite small positive  cosmological
or quintessence energy \ci{vilenkin}.
 Of course,
 there is a fine tuning problem such that the difference of two quantities at
the  high energy scale  gives something of the order of today's
energy $\Lm_c\sim 10^{-12} GeV$ and that such a cancellation must
take place during nucleosynthesis and today.

However,   the evolution of the scalar field with negative potential
will necessarily   give a   region with a small cosmological constant,
since for $\rho_\phi$ negative we have
$|\rho_\phi|<\rho_b$ regardless of the functional form
of the potential $V$ and the fact that $\wp<-1$ which makes
$|\dot\rho_\phi|$ grow faster than $\rho_b$.
The evolution is such that $\rho_b+\rho_\phi\rightarrow 0$
for any potential $V$ or energy density $\rho_b$. So it is
an attractor solution.

During all the time that $\rho_\phi<0$ we have $|\rho_\phi| < \rho_b$,
i.e. it will get smaller than the sum of all other energy densities
for a long period of time and regardless of the functional form
of the potential $V$.

\section{Energy Conditions}

 The weak energy condition "WEC" states that the total energy density $\rho$
and pressure $p$ obey the inequalities
\bea\la{wec}
\rho+p=\rho(1+w) &\geq& 0\non\\
\rho &\geq&  0.
\eea
where we have used   the equation of state $p=w\rho$. If WEC is
supplemented with $\rho \geq |p|$   then it is called the dominant
energy condition "DEC". The argument for requiring $\rho \geq |p|$
is because for   a constant equation of state $w=p/\rho$, $w$
gives the speed of sound and it should not exceed the speed of
light  $c=1$, i.e. $|w|\leq 1$. However, if $w$ is not constant
then we can have regions with $|w|>1$ and  $|dp/d\rho|>1$ without
violating causality. A full discussion on the sound speed, signal
speed  and causality for scalar fields is given in \ci{ax.vuc}.

In a flat universe, the second
equation in (\ref{wec}) can be  understood also from the positivity
of the Friedman equation $3 H^2=\rho$  and the first
equation is obeyed by matter ($w=0$), radiation ($w=1/3$), cosmological
constant ($w=-1$) and
scalar fields  ($\rho+p=2E_k=\dot \phi^2$).

 Here we will concentrate on the cosmology behavior of a canonically normalized
scalar  field $\phi$, $E_k=\dot \phi^2/2$,  with a  potential  that has a negative minimum
$V(\phi)|_{min}<0$  in a flat universe. Having a scalar
field with a negative
potential will lead dynamically to a region with $|\wp\equiv
p_\phi/\rho_\phi|>1$.

In a flat universe we have
\be\la{h}
H^2=\fr{1}{3}\rho
\ee
with   the reduced Planck mass $m^2_p=1/8\pi G\equiv 1$ and with
$\rho=\Sigma_A\rho_A$, where $A$ stands for all kinds
of energy densities, matter, radiation,
cosmological constant or scalar fields. It is
usually stated that a flat universe  expands
forever and it is infinite but this conclusion is not completely correct. It is
based on the assumption that all energy densities are positive
$\rho_A > 0$ at finite time. However, if the contribution from some kind of
energy has $\rho_A < 0$, then it is possible to have $3H^2=\rho=0$
at finite time and the size of the universe will start to decrease
afterwards. The behavior is very similar as for a closed universe with
positive curvature
since it will reach a maximum size and then will start contracting
ending in a big crunch.

To see this, let us study the  cosmological evolution of $\phi$
with an arbitrary potential $V(\phi)$. It can be  determined from
a system of differential equations describing a spatially flat
Friedmann--Robertson--Walker universe,
\bea\la{Hro}
\dot H &=& -\frac{1}{2}(\rho + p), \\
\dot \rho &=& -3H(\rho+p)\non
\eea
where $H\equiv \dot a/a$ is the Hubble parameter and  dot refers
 to derivative w.r.t. time.

The weak energy condition
together with eqs.(\ref{Hro}) imply that $H$ is an non increasing
quantity, $\dot H \leq 0$.  Since the scale factor is always
non-negative, $a \geq 0$, the maximum size of the universe will be given
at $H(t_0) = 0$ with $a(t) \leq a(t_0)$
and afterwards the scale factor will start to shrink
until it arrives at a big crunch with $a\rightarrow 0$.
For this kind of behavior, it is required that the total energy
density vanishes at a finite time. For $H>0$ the energy density $\rho$
decreases with time and it reaches $\rho(t_0)=0$ at $H(t_0)=0$. If this happens
at a finite time $t_0$ then the evolution in eq.(\ref{Hro})  leads to a negative $H$
and a increasing $\rho$ for $t>t_0$.

A third possibility is that of a
 phantom energy. The phantom energy does not comply with the WEC since it
has $\rho_{ph}>0$ but $  \rho_{ph}+p_{ph}=\rho_{ph}(1+w_{ph})=2E_k<0$, i.e.
 the scalar field has  non-canonical (negative) kinetic term.
 From eqs.(\ref{Hro}) we can  see that if the phantom field dominates then $\dot H$
is positive
giving an increasing $H$ and  $\rho_{ph}$. The fact that phantom physics
gives a big smash can be understand from the solution of the scale factor
 valid for $w_{ph}$   constant and  when
 the phantom field starts to dominate and after a  period of matter domination.
The solution is
$a(t)=a(t_m)[-w_{ph}+(1+w_{ph})(t/t_m)]^{2/3(1+w_{ph})}$
\ci{cald}. It is clear that $a\rightarrow \infty$ at the finite
time $t=t_m w_{ph}/(1+w_{ph})$ since the exponent of $a(t)$ is
negative ($1+w_{ph}<0$) and the term in brackets vanishes.
 As mentioned above, here we will not
work with this kind of energy density since we are only considering
standard canonically  normalized scalar fields.

It is clear, from the Friedman and WEC equations, that the total energy
density must be positive semi-definite   but we can have a sector with
negative energy
densities, say $\rho_i$, without contradicting the WEC (eq.(\ref{wec}))
as long as the total energy $\rho=\Sigma_A \rho_A \geq 0$. However,
even for this type of energy density, which we assume comes from
a canonical normalized scalar field  we have
\be\la{wec1}
\rho_\phi+p_\phi=\rho_\phi(1+\wp)=2E_k \geq 0
\ee
with $\rho_\phi=E_k+V, p_\phi=E_k-V$. This inequality is clearly also
satisfied for  matter, radiation or cosmological constant.  Therefore,
$\rho_i+p_i\geq 0$ holds for all classes of
fluids we are here considering.  However, eqs.(\ref{wec1}) allows
for a negative energy density as  along as $w_\phi < -1$, i.e. we
can have
\be\la{w1}
\rho_\phi < 0 \hspace{5mm} \Leftrightarrow \hspace{5mm} w_\phi < -1
\ee
as long as the total $\rho\geq 0 $ with the total $w=p/\rho >-1$.
We will also have for $\rho_\phi>0$
\be\la{w1b}
E_k>-V>0 \hspace{5mm} \Leftrightarrow \hspace{5mm} w_\phi >1.
\ee

The fact that we have $\wp<-1$ is not a problem with causality because
it is only in the case of $\wp$ constant that $\wp$ can be interpreted as
the  sound speed of the  fluid in which case it must be
smaller than the speed of light.

\section{Time Inversion Symmetry and Dynamics}

Let us now study under which conditions do we have
a time inversion symmetry around the point
$t_0$, where $H(t_0)=0$. The dynamical equations
are
\bea\la{H}
H^2(t)&=&\frac{1}{3}\sum_A \rho_A (t)\\\la{dH}
\dot H(t) &=& -\frac{1}{2}\sum_A(\rho_A(t) + p_A(t)), \\
\la{dr}
\dot \rho_A(t) &=& -3H(t)(\rho_A(t)+p_A(t))
\eea
where $A$ stands for all kind of fluids, matter,
radiation, cosmological constant or scalar fields.

To have an inversion symmetry it is sufficient
to have $\rho_A (t)=\rho_A (t')$
and $p_A (t) = p_A ( t')$ (or equivalently
$\rho_A$ and $\rho_A(1+w_A(t))=\rho_A (t)+p_A ( t)$ symmetric)
for all fluids  for $t=t_0-\Delta t$ and $t'=t_0+\Delta t$
(for $t_0=0$ we would have a transformation
$ t\rightarrow t'=-t$).
For fluids with constant $-1\leq w_A \leq 1$, as matter, radiation
or cosmological constant, the time symmetry is
always satisfied. Therefore, the big crunch is just
as problematic as the big bang and the solution
  could be expected to be the same for both singularities.

If we have scalar fields, then a time inversion symmetry
is only possible  if $\dot\phi^2(t)=\dot\phi^2(t')$ for all $t$,
since $\rho_\phi+p_\phi=\rho_\phi(1+w_\phi)=\dot \phi^2$. This
will be satisfied for $\dot \phi(t)=\pm \dot\phi (t')$. If
\be\la{cond1}
\dot\phi(t)=-\dot\phi(t')\hspace{5mm} \Leftrightarrow \hspace{5mm}
\dot\phi(t_0)=0 \hspace{5mm}\Leftrightarrow \hspace{5mm}\phi(t)=\phi(t')
\ee
Clearly the time inversion symmetry would be, in this case, valid for
all kinds of potentials $V(\phi)$. If $\dot \phi(t)=\pm \dot\phi (t')$
is satisfied locally around $t_0$ the dynamics ensures that
it is valid for all $t$.
However, if
$\dot \phi(t_0) \neq 0$, then a symmetric solution
is only possible if the potential $V(\phi)$ is
symmetric around $\phi(t_0)$. This can be seen
from the equation of motion of $\phi$
\be\la{phi}
\ddot \phi +3H\dot \phi+V'=0
\ee
with $V'=dV/d\phi$ and at $t_0$ one has $H(t_0)=0$
and for $\dot\phi(t_0)\neq 0$ a symmetric solution
requires
\be\la{cond2}
\dot\phi(t)=\dot\phi(t')   \hspace{5mm} \Rightarrow
 \hspace{5mm}
\ddot\phi(t_0)=0
\hspace{5mm}\Leftrightarrow \hspace{5mm}V'(t_0)=0.
\ee
We conclude that arbitrary initial conditions at $t_0$ for
a scalar field will, in general,  break the time inversion
symmetry. Only a symmetric initial condition given at $t_0$
with $\dot\phi(t_0)=0$  will respect the time symmetry independently
of the potential $V$, or for a symmetric potential around $\phi(t_0)$
and arbitrary initial condition for $\dot\phi$.

As a consequence  of the time inversion symmetry
the final time is $t_f\simeq 2t_0-t_i$, with $t_i$
the initial time which could be set to  zero, and  it does not make
sense to take $t_f\rightarrow \infty$. There will be
a big crunch only if there was a big bang singularity.
The minimum size of the scale factor is then at $a(t_f)=a(t_i)$,
and the energy densities and Hubble parameter will end up at the same
value as they started $\rho_A(t_i)=\rho_A(t_f),\;H(t_i)=-H(t_f)$.
So the dynamics for  $t>t_0$ is just the same as
the dynamics for $t<t_0$. In \ci{linde} it was argued that
negative potential are as dangerous as unbounded potentials
and that the kinetic terms dominates at large times. This
is true but the "reasonable" position is to cut off the
time at $t_f=2t_0-t_i$.

\subsection{Solution with constant $w$}

It is useful to study first the evolution of
fluids with constant $w$ and later introduce
a scalar field with arbitrary potential
varying $w$.

The main features can be obtained
from the $w_b$ constant example. From eq.(\ref{dr})
an arbitrary energy
density evolves as
\be\la{rb}
\rho_b(t)=\rho_{b i} \le (\fr{a(t)}{a_i}\ri)^{-3(1+w_b)}
\ee
with a constant equation of state parameter $-1\leq w_b \leq 1$
and  initial energy density $\rho_{b i}>0 $.

As long as we have an expanding universe $H>0$, i.e. $a(t)>a_i(t_i)$ for
$t>t_i $,
and with  $1+w_b>0$ we have a decreasing amplitude in time
of $\rho_b$ with $\rho_b \rightarrow 0$ at late times.
This kind of  behavior is the usual one for radiation or matter, with $w_b=1/3,0$
respectively.

However, as discussed above, we can have $H < 0$.
 In this case we have a decreasing scale factor
$a(t)< a_i(t_i)\;(t>t_i)
$ and $\rho_b$ increases with time (decreasing
$a(t)$). It is easy to see that we have a symmetric
$\rho_b$ under the transformation $t=t_0-\Delta t\rightarrow t'=t_0+\Delta t$
with $H(t_0)=0$ if $a(t)=a(t')$.
 Eq.(\ref{rb}) can be rewritten as
\be\la{rb2}
\rho_b(t)=\rho_{b 0} \le (\fr{a(t)}{a(t_0)}\ri)^{-3(1+w_b)}
\ee
with $a(t)<a_0(t_0)$ and    valid for all $t$. The energy density
$\rho_{b 0}$ gives the minimum value of $\rho_b$.

Let us now include   a cosmological
constant with negative energy
density $\rho_\Lm=-p_\Lm=-\Lm<0$ and a fluid with constant
$|w_b=p_b/\rho_b|<1$.

In this case we will have a symmetric
solution around $t_0$.

The equations to be solve are  $\dot\rho_\Lm=0$
and $\rho_b=-3H\rho_b \ga_b$, with $\ga_b\equiv
1+w_b=cte$.
Using $H=\pm \sqrt{(\rho_b-\Lm)/3}$,
with the plus sign for $t<t_0$ and the minus sign for
$t> t_0$, we can integrate eq.(\ref{dr}) from $t$ to $t_0$
giving
\be
\int \fr{d \rho_b}{\rho_b\sqrt{\rho_b-\Lm}}=-3\ga_b\mid t_0-t \mid
\ee
and
\be\la{rb3}
\rho_b (t)=\Lm \;\;Tan
\le[\fr{\ga_b\sqrt{3\Lm}}{2}\;\;( t_0-t) \ri]^2+\Lm
\ee
with
\be\la{hh}
H(t)= \sqrt{\fr{\Lm}{3}}\;\; Tan\le[\fr{\ga_b\sqrt{3\Lm}}{2}\;\;(t_0-t) \ri]
\ee
were we have taken $\rho_b(t_0)=\Lm$ (i.e. $H(t_0)=0$). The Hubble
parameter $H(t)$ in eq.(\ref{hh}) is positive for $t<t_0$ and it is negative
for $t>t_0$.
It is easy to see that the solution  (\ref{rb3})
and (\ref{hh}) are symmetric under time inversion
$t_0-t \rightarrow t_0+t$
and $\rho_b, H$ blow up at $\fr{\sqrt{3\Lm}\ga_b}{2}(t_0-t)=\pi/2$
since $Tan[\pi/2]=\infty$.
The quantities $\rho_b, H^2$ start at a given initial value and they
decrease as long as $t<t_0$ reaching its minimum value
at $t_0$ with $\rho_b(t_0)=\Lm$ and $H(t_0)=0$. For $t>t_0$ we
have a negative $H$ and $\rho_b$ starts to increase. We would
take the final value of $t$ as $t_f-t_0=t_0-t_i$ and $\rho_b(t_f)=\rho_b(t_i)$.
We can also see that $\rho$ never becomes negative nor $\rho_b$ smaller
than $\Lm$.

\subsection{Scalar field Dynamics}

We will now study the evolution of scalar field with
a negative minimum $V|_{min}<0$ potential with energy density
$\rp=E_k+V$ and pressure $p_\phi=E_k-V$. This situation
is completely equivalent as to having a semi-positive
definite potential
in the presence
of a negative energy density.

Considering a scalar field with negative potential but with no
other type of energy densities, then WEC requires
$ E_k +V \geq 0 $ and therefore $\wp$ is bounded from below $\wp\geq -1$
at all times. In this case we cannot arrive at the situation where
we have $\wp<-1$ and $\rho_\phi <0$, even though we have a negative
 $V$ but there will be regions with $\wp>1$.

For  two (or more) sectors which are
only connected via gravity so that $\rho=\rho_b+\rho_\phi,\;
p=p_b+p_\phi$ with $\dot\rho_A=-3H(\rho_A+p_A)$ and
equation of state $p_A=w_A\rho_A$ for $A=b,\phi$.
In this case the weak energy
principle is $\rho=\rho_b+\rho_\phi\geq 0$  and $\rho_A+p_A=\rho_A(1+w_A)\geq 0$
for each component separately   $A=b,\phi$.

To have an idea on how the evolution of the different energy densities
behave  in the presence of a scalar field with
negative potential  we  solve eq.(\ref{dr})
for $\rho_\phi$ with  constant $\wp$ and consider the different
regions. Of course, $\wp$ is far from being a constant but
the generic features are well described by taking
$\wp$ constant with different values. We will later allow
for a dynamical non constant $\wp$.

An arbitrary energy density evolves as in eq.(\ref{rb}).
\be\la{rp}
\rp(t)=\rho_{\phi i} \le (\fr{a(t)}{a_i}\ri)^{-3(1+\wp)}.
\ee
where  $\rho_{\phi i},\; \wp$ take
different values depending in which region they
are taken. As long as we have an expanding universe $H>0$, i.e. $a(t)>a_i$,
and we can distinguish two different cases
$\rho_{\phi i}$ positive or negative.
For $\rho_{\phi i}>0$ we have
$1+\wp>0$ and a decreasing amplitude
of $\rp$. However, as discussed above, we can have $\rho_{\phi} \leq 0$
with total $\rho\geq 0$. In this case, from eq.(\ref{w1})  we
necessarily have $\wp<-1$. Still in the region with  $H>0$ we
have a decreasing $\rho_\phi$ but since it is negative $|\rho_\phi|$
grows in time, i.e. it becomes more and more negative. In the mean
time the energy density of all other sectors (with $\rho_b>0$)
are also decreasing (in absolute terms), so we will necessarily
reach a point when  the total energy density vanishes
$\rho=\rho_b+\rho_\phi=0,\;H=0$.

During all the time that $\rho_\phi<0$ we have $|\rho_\phi| < \rho_b$,
i.e. it is smaller than the sum of all other energy densities
for a long period of time and regardless of the functional form
of the potential $V$ and the fact that $\wp<-1$.

Since $H$ is always decreasing, after $H(t_0)=0$
we will have a contracting universe with  $H<0$, i.e. $a(t)<a(t_0)$.
At this time  $\rho_\phi$, which is still negative,
will start growing with a $\wp<-1$ and become less
negative. Eventually,
$\rho_\phi$ will become positive and  $\wp>-1$.
The energy
density will continue to grow with decreasing scale factor $a(t)$.
If we have a symmetric solution for the scalar fields
(see eqs.(\ref{cond1}) and (\ref{cond2})) than
$a(t_f) \rightarrow a_i(t_i) $ and $\rp \rightarrow \rho_{\phi i}$
where $a_i,\rho_{\phi i}$ are the initial values.

Until now, we have shown the generic evolution of $\rho_\phi$
considering $\wp$ constant and taking different values depending on
the region.

In \ci{linde} it was shown that a scalar field with negative
potential its kinetic energy will dominate over the scalar potential $V$
but this result  depends on the potential.
Let us now see under which conditions will the kinetic term dominate
the potential. In this approximation we can
take in  eq.(\ref{phi}) $V'\ll 3H\dot\phi$, i.e.
\be\la{ddphi}
\ddot \phi +3H\dot\phi=0.
\ee

In order to analytically solve eq.(\ref{ddphi}) we
can take, without loss of generality,
the scalar potential $V(\phi)=V_1(\phi)
+V_{min}$, with $V_{min}\equiv-\Lm$ the (constant)
value of $V$  at the minimum and  $V_1\geq 0$.
In the case that $\rho_b+V_{min}=\rho_b-\Lm$ dominates over $V_1$
we can use in eq.(\ref{ddphi})
$H$ given by eq.(\ref{hh}).  Doing so,
we have $d\varphi/\varphi=-3Hdt$ with $\varphi\equiv \dot\phi$
and integrating the r.h.s.  we find
\bea\la{intH}
\int_{t}^{t_0} \fr{d \varphi}{\varphi}&=&
-\int_{t}^{t_0} 3Hdt\non\\
Log\le[\fr{\varphi(t_0)}{\varphi(t)}\ri]&=&
-\fr{2}{\ga_b} Cos\le(\fr{\sqrt{3\Lm}\ga_b}{2} (t_0-t) \ri)
\eea
giving
\be\la{dphi}
\dot \phi(t)= \dot \phi_0 \le[ Cos\le(\fr{\ga_b\sqrt{3\Lm}}{2} (t_0-t) \ri)\ri]
^{\fr{2}{\ga_b}}.
\ee
For a matter background $\ga_b=1+w_b=1$
the exponent in eq.(\ref{dphi}) is 2 and we can solve it
for $\phi=\int \dot\phi dt$ giving
\be\la{phi2}
\phi(t)=\phi_0+\fr{t-t_0}{2}+\fr{1}{2\sqrt{3\Lm}}\;\; Sin\le(\sqrt{3\Lm}
(t_0-t)\ri).
\ee
We see from eqs.(\ref{dphi}) and (\ref{phi2}) that the kinetic
term and the scalar field (or potential $V=c\phi^n$)
have an oscillating behavior and neither
of them diverges  large at large $t$. Of course, this result
is only approximated since it  has
been obtained under the hypothesis that the barotropic fluid
dominates so that we can use $H$ of eq.(\ref{hh}).

In the other limiting case where $\ddot\phi \ll 3H\dot\phi=-V'$
with a potential $V=c\phi^n$, we
have $d\phi/V'=d\phi/cn\phi^{n-1}=-dt/3H$ and
\be\la{phi3}
\phi(t_f)^{2-n}-\phi(t_i)^{2-n}=\fr{2n(n-2)c}{3\Lm}
Log\le[\fr{Sin\le(\fr{\sqrt{3\Lm}}{2}(t_0-t_f)\ri)}{
Sin\le(\fr{\sqrt{3\Lm}}{2}(t_0-t_i)\ri)}\ri]
\ee
for all $t$ (larger or smaller than $t_0$).
In this case there is a  divergent behavior at $t\rightarrow t_0$
where $Log(Sin[t\rightarrow t_0]) \rightarrow -\infty$
and $|\phi(t)^{2-n}| \rightarrow \infty$. For $n>2$ one has
$\phi(t) \rightarrow 0$ while for $n<2$
the value is $\phi(t) \rightarrow \infty$.

\section{Effective w}

For an arbitrary potential $V$, the quantity $|w_\phi=(E_k-V)/(E_k+V)|<1$
 at all times if $V$ is  a non-negative potential since
$E_k+V > E_k-V$. However for negative potentials we have regions
with $|\wp|>1$, since in this case  the terms in the numerator have
the same  sign, when $V$ is negative, while in the  denominator they
have opposite sign. The parameter   $w_\phi$ will then be greater than one for
$\rho_\phi=E_k+V>0$ and smaller than -1 for $\rho_\phi<0$
as given by eqs.(\ref{w1}) and (\ref{w1b}).

As stated in the introduction, the SN1a and CMB data do not ruled
out an energy density with $w<-1$. We have seen so far that
in order to have $\wp<-1$, for a canonical scalar field, we also need
to have $\rho_\phi<0$. So, what is the effect of having a negative energy density
with $\wp<-1$ in the total equation of state $w=p/\rho$?

Let as consider an energy density given by
$\rho=\rho_1+\rho_2$ with the conditions
$\rho>0,\rho_1>0,\rho_2<0$ and equations of state
$p=w\rho,\,p=p_1+p_2$ and $ p_i=w_i\rho_i,\,i=1,2$. As we have
seen previously,  $w_2$ must necessarily be smaller than -1 and
we take $-1<w_1<1$. The energy density $\rho_1$ can be either matter,
radiation, cosmological constant or a scalar field with non negative potential (e.g.
quintessence $-1\leq w_1 \leq 1$). The equation of state parameter for
the total $\rho$ is
\be
w=\fr{p}{\rho}=\fr{w_1\rho_1+w_2\rho_2}{\rho}
\ee
Since $\rho_1> \rho=\rho_1+\rho_2>0$ and $w_2\rho_2>0$ (both are
negative) we have
\be
 w>w_1\rho_1/\rho > w_1
\ee
regardless of the values of $\rho_1,\rho_2$. Notice that the value
of $w_2$ is smaller than -1 but the final $w$ is not only greater
than -1  but it is greater than $w_1$. This means that a negative
potential will increase  the value $w_1$ of any energy
density  component  including that of a scalar field with positive energy
density or that of  a cosmological constant with
$\rho_1=cte$ and $w_1=-1$.

A scalar field with a negative minimum potential
 can always be  written as
$V=V_1+V_{min}$ with $V_1\geq 0$ and $V_{min}<0$ constant. The
minimum of the potential is at $V_1=0$ and $V=V_{min}<0$. This separation
of the potential for a scalar field  is entirely equivalent
as to having a scalar field with non negative potential $V_1$
and kinetic energy $E_k=\dot \phi^2/2$
in the presence of a negative cosmological constant given by
$\rho_\Lm=V_{m
in}=-p_\Lm<0$. The equation of state would be
\be
|w_1|=|\fr{p_1}{\rho_1}|=|\fr{E_k-V_1}{E_k+V_1}|<1 \hspace{1cm}
 w_\Lm=\fr{p_\Lm}{\rho_\Lm}=-1.
\ee
The  original equation of state $\wp=(E_k-V)/(E_k+V)=(E_k-(V_1+V_{min}))/(E_k+V_1+V_{min})
$ can be expressed in terms of $w_1$ and $w_\Lm=-1$ as
\be
\wp=w_1-(1+w_1)\fr{\rho_\Lm}{\rho_1+\rho_\Lm}
\ee
and if $\rho_1+\rho_\Lm>0$ we have $\wp<-1$
since $\rho_\Lm/(\rho_1+\rho_\Lm)<-1$ (the terms in the denominator
have opposite signs)   while for  $\rho_1+\rho_\Lm<0$  we get $\wp>1$
since $\rho_\Lm/(\rho_1+\rho_\Lm)> 1$ and we
recover eqs.(\ref{w1}) and (\ref{w1b}).

We conclude that a scalar field with negative potential
is equivalent to a positive semi-definite potential
in the presence of a negative energy density cosmological
constant.

\section{Models with Negative Potentials}

Let us now analyze the cosmological evolution
of scalar fields for different negative potentials.

In the first case, we consider a potential with
a simple mass term $V=m^2\phi^2
+V_2$ with $V_2=-\Lm<0$ in the presence of matter.
The evolution is shown in fig.\ref{fig1}.

\begin{figure}[htp!]
\begin{center}
\includegraphics[width=7cm]{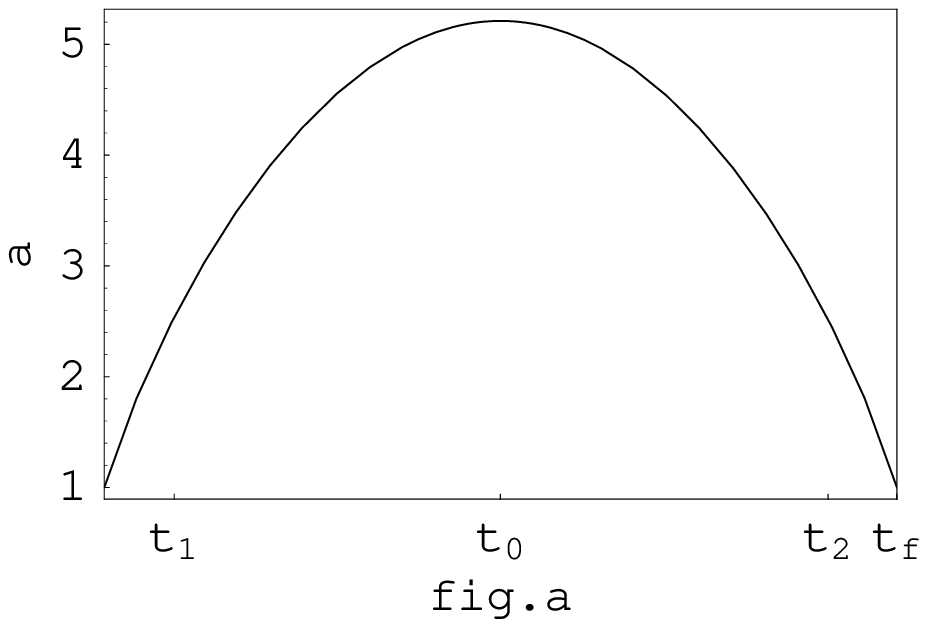}
\includegraphics[width=7cm]{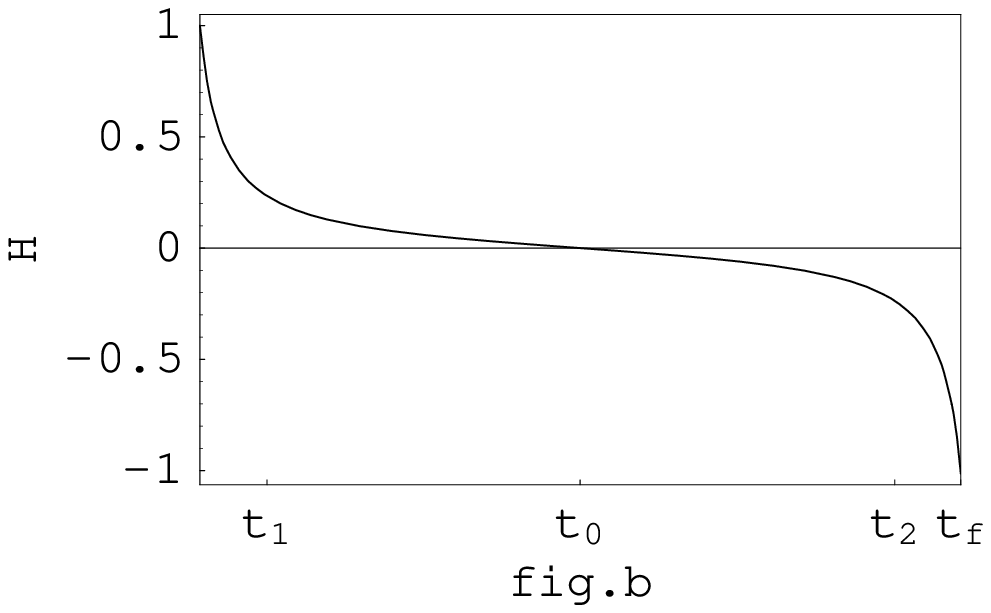}
\includegraphics[width=7cm]{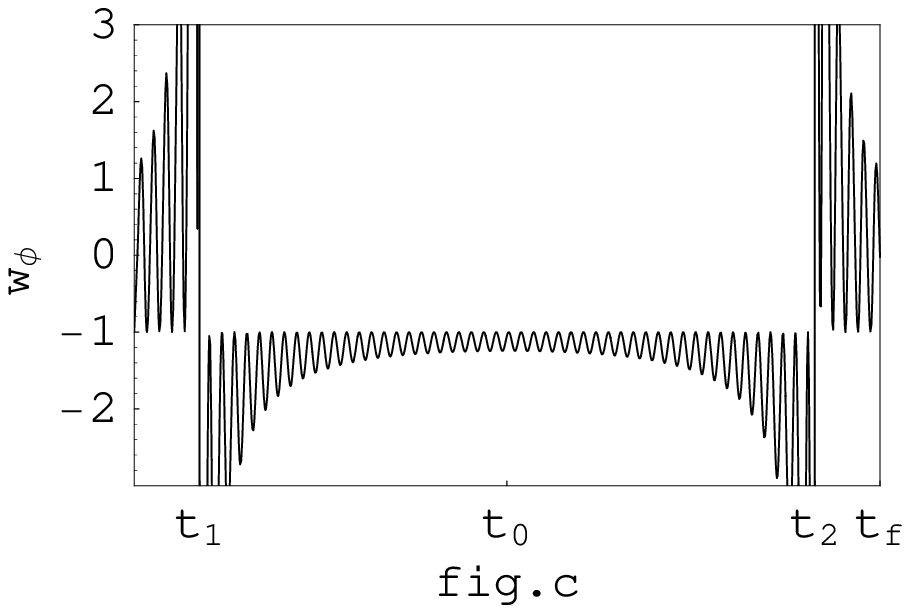}
\includegraphics[width=7cm]{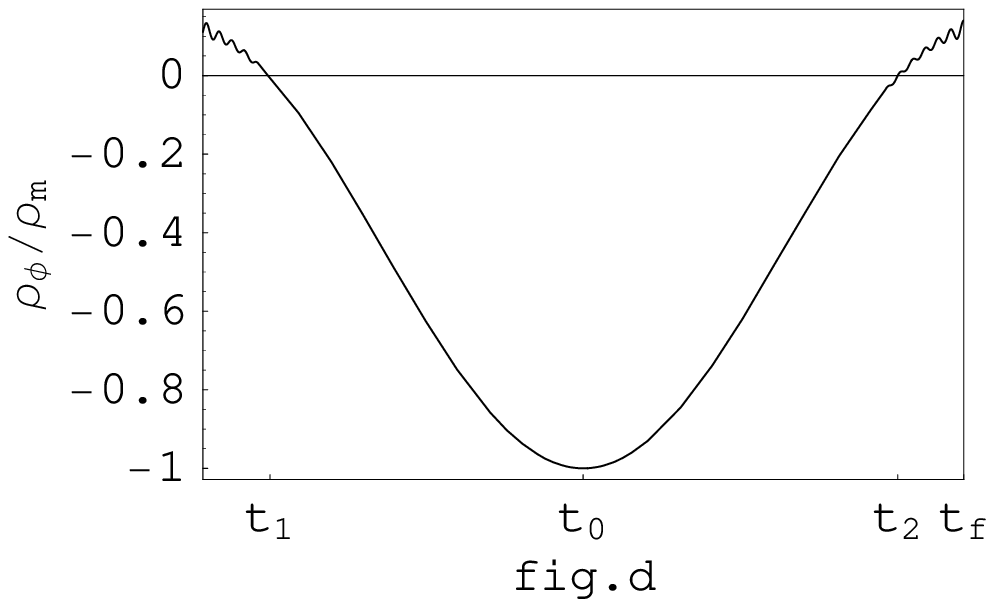}
\end{center}
\caption{\small{ We show the evolution for
a scalar field with potential $V=m^2\phi^2+V_2,\;V_2=-\Lm<0$.
The field oscillates  around $\phi=0$ and we have
$\wp<-1$ for $t_1<t<t_2$, where $t_0$, $t_1$ and $t_2$ are
defined
by  $H(t_0)=0, \rp(t_1)=\rp(t_2)=0$.
Even though at $t_0$, we have $\phi_0\neq 0$
and $\dot\phi_0\neq 0$, i.e. no symmetric
initial conditions, the solution is
almost symmetric around $t_0$ since
$|\phi_0|\ll 1,|\dot\phi_0|\ll 1$.
}}
\la{fig1}
\end{figure}

\begin{figure}[htp!]
\begin{center}
\includegraphics[width=7cm]{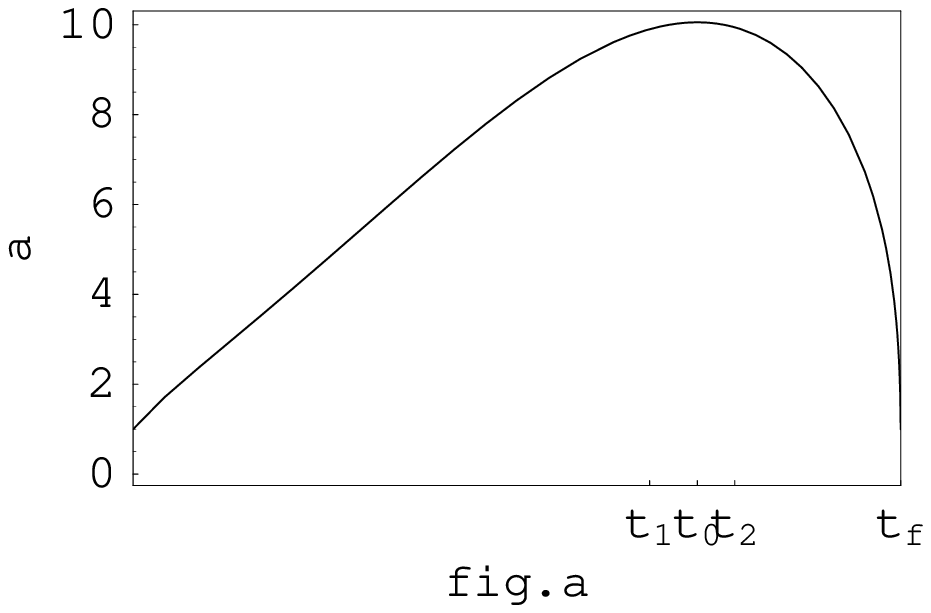}
\includegraphics[width=7cm]{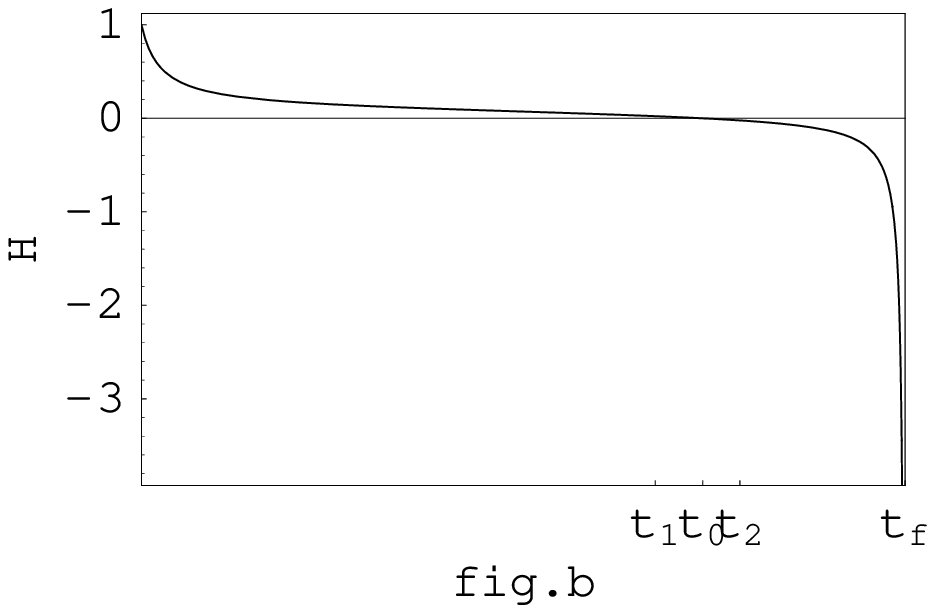}
\includegraphics[width=7cm]{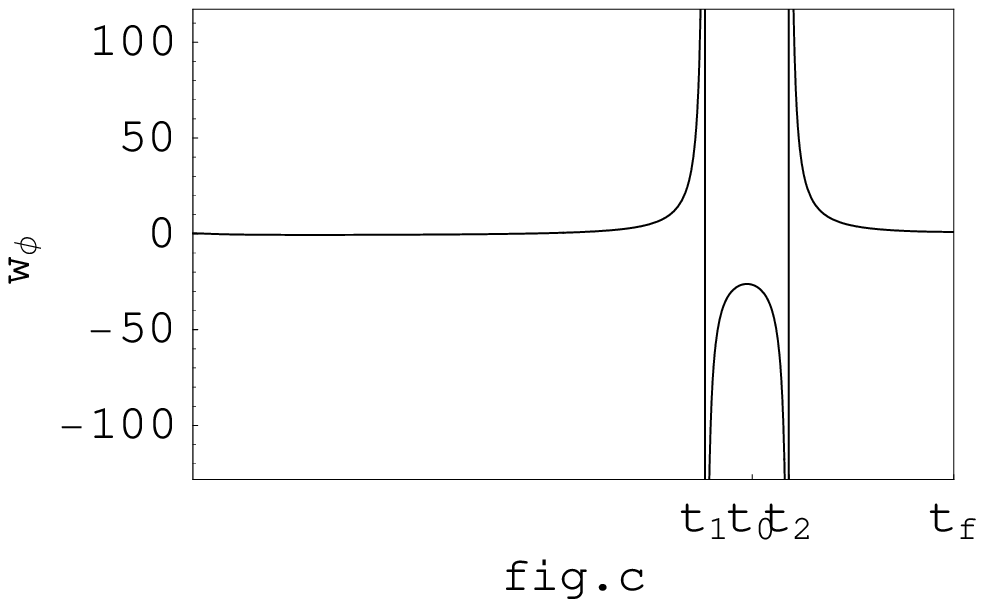}
\includegraphics[width=7cm]{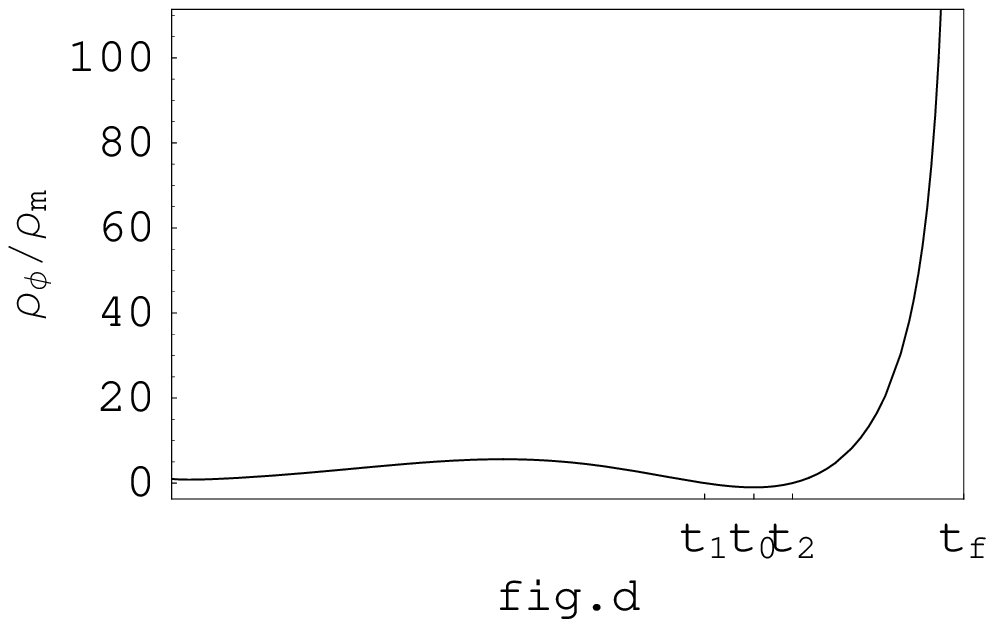}
\includegraphics[width=7cm]{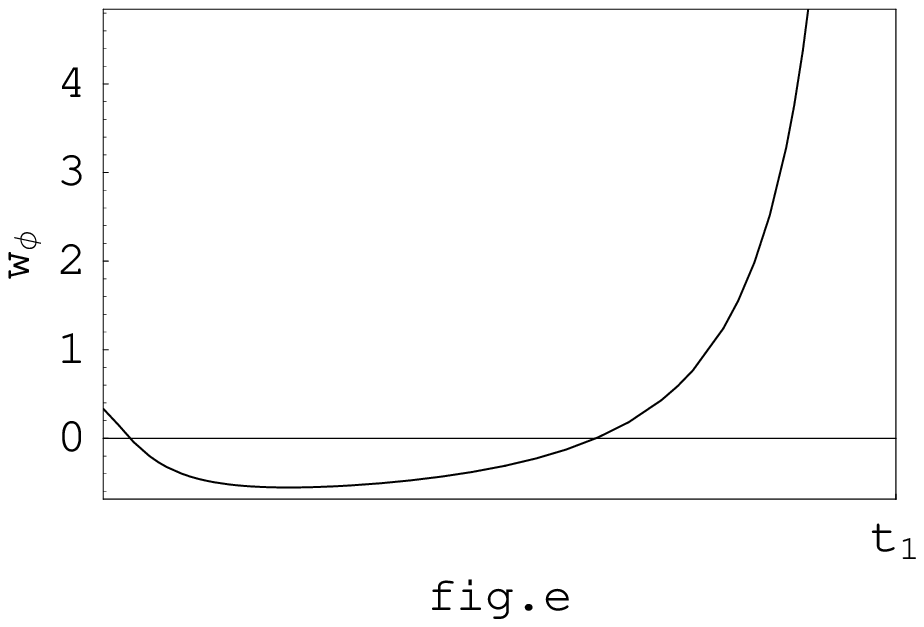}
\end{center}
\caption{\small{ We show the evolution for a scalar potential
$V=m\phi^{-1}-\Lm$ with initial conditions at $\dot\phi(t_0)\neq
0$ and the times $t_1<t_0<t_2$ defined by  $H(t_0)=0,
\rp(t_1)=\rp(t_2)=0$. Notice that the model has no time inversion
symmetry around $t_0$ since $\dot\phi(t_0)\neq 0$ and the
potential is not symmetric $\wp<-1$ for $t_1<t<t_2$. From fig.e we
see that there is a region with $<-1\wp<0$ where the universe
could be in a accelerated epoch. }} \la{fig2}
\end{figure}

\begin{figure}[htp!]
\begin{center}
\includegraphics[width=7cm]{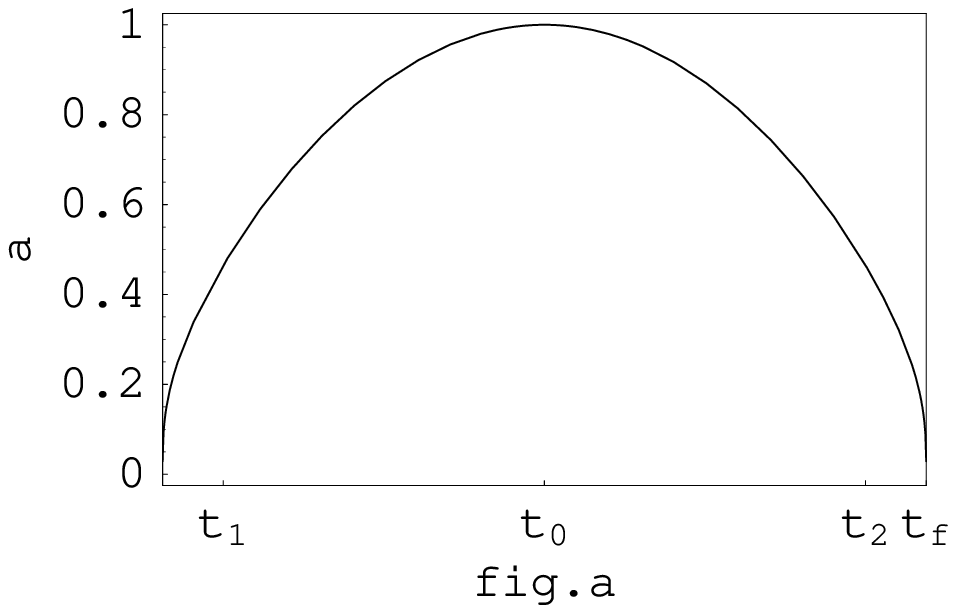}
\includegraphics[width=7cm]{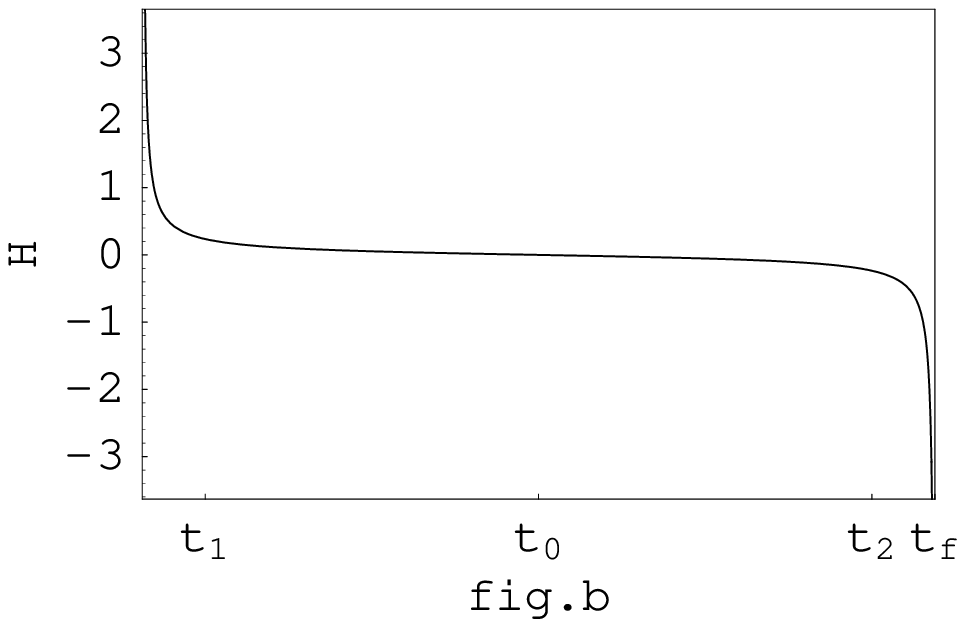}
\includegraphics[width=7cm]{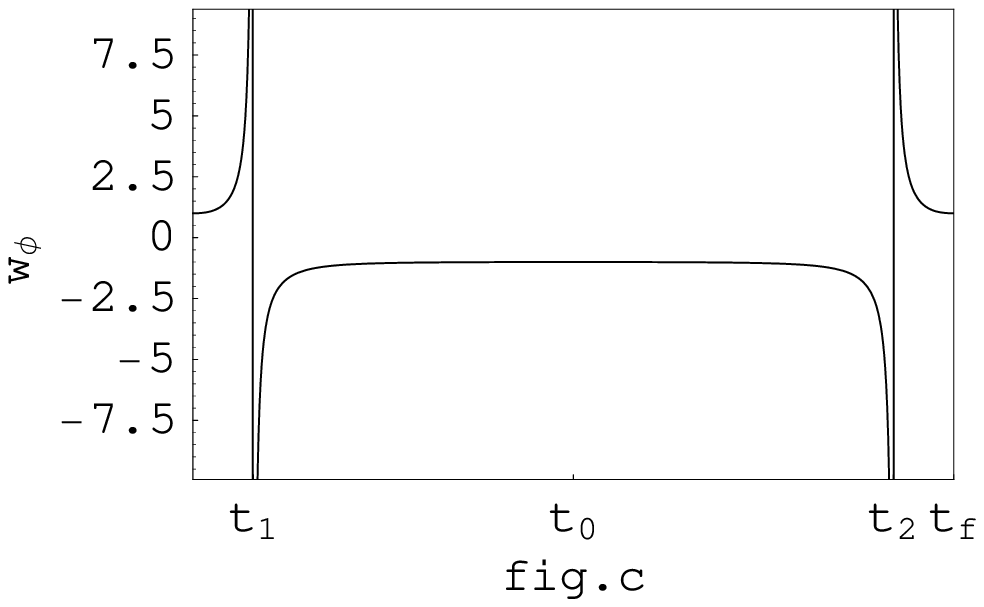}
\includegraphics[width=7cm]{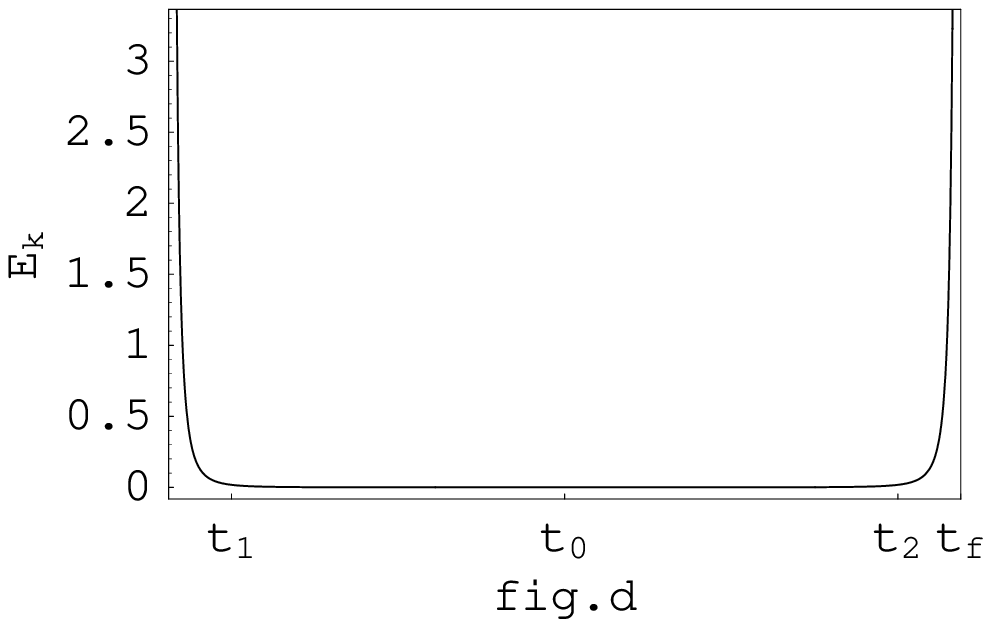}
\includegraphics[width=7cm]{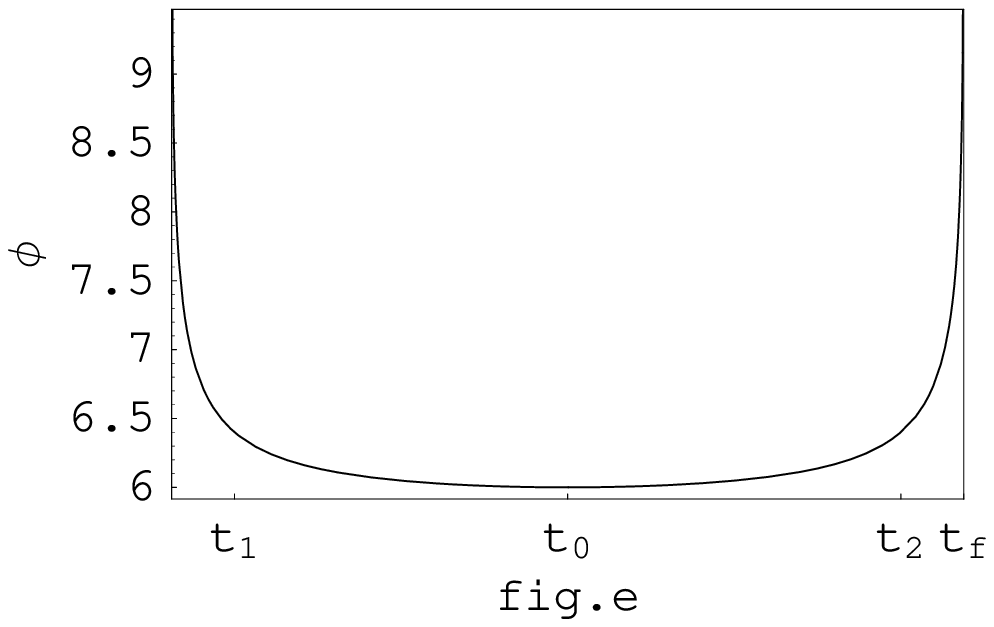}
\end{center}
\caption{\small{ We show the evolution for
a scalar potential $V=m\phi^{-1}-\Lm$ with
initial conditions at $\dot\phi(t_0)= 0$
and the times $t_1<t_0<t_2$ defined
by  $H(t_0)=0, \rp(t_1)=\rp(t_2)=0$.
Notice that the model is symmetric
around $t_0$ since $\dot\phi (t_0)= 0$
even though the potential is not symmetric.
Once again, $\wp<-1$ for $t_1<t<t_2$.
}}
\la{fig3}
\end{figure}

As given by solving eqs.(\ref{H}) the field rolls down the
potential and oscillates around it. For $t<t_0$ we have an
expanding universe with $H>0$ while for $t>t_0$ we have a
contracting universe with $H<0$. While $\rho_\phi \geq 0$ and
$V\geq 0$ we have $-1\leq \wp \leq 1$ and as soon as $V<0$ we have
$\wp >1$.
 At $t_1$ and $t_2$  we have $\rp(t_1)=\rp(t_2)=0$ and we see that
for $t_1<t<t_2$ the equation of state parameter oscillates with amplitude
$\wp<-1$ and the energy density is $\rp(t)<0$.
For $t<t_0$ the energy density $\rp$ decreases but
it increases in magnitude for $t_1<t<t_0$. We have an almost
symmetric phase around $t_0$ even though $\dot\phi(t_0) \neq 0$
this is because the field is oscillating around the minimum
with decreasing amplitude for $t\leq t_0$ and the potential
itself is symmetric around the $\phi=0$.

For $V>0$ we can have an accelerating period if $\phi$ is large
enough to satisfy the slow roll conditions. At the end of the slow
roll the field will oscillate around the (negative) minimum where
$w_\phi < 0$. The universe will recollapse when $\rp$ equals the
background energy density and $H=0$. This could well happen in the
future. However, in order to have $\phi$ as the quintessence field
its appearance would have to be fine tuned to give
$\Omega_\phi\simeq 0.7$ at present time and not to accelerate the
universe at times earlier than $Log[a_o/a] > 1$.

As a second example we consider an inverse power
potential with $V=m\phi^{-1}+V_2$, with $V_2<0$
constant in the presence of matter with $w=1$.
In this case the potential is clearly
not symmetric for any value of $\phi$. So, we
would expect a symmetric solution only
if $\dot\phi(t_0)=0$. In fig.2 and fig.3
we see the evolution of the scale factor $a$, the Hubble
parameter $H$ and the equation of state $\wp$ for
different initial conditions. In fig.2 we have
$\dot\phi(t_0) \neq 0$ and we can see that there
is no symmetry for $t<t_0$ and $t>t_0$. However, the
generic behavior of the energy densities $\rm,\rp$
is still valid, i.e. they decrease for $t<t_0$
and they increase for $t>t_0$. The equation
of state parameter $\wp<-1$ for $t_1<t<t_2$.

Notice that for $t<t_1$ the value of $\wp$ is
negative, i.e. $-1< \wp<0$. In the absence
of the negative term, we know that inverse power
potentials (as $V=m/\phi$) lead to an accelerating
epoch and to $\Omp \rightarrow 1$, dominating
the universe at late times and the acceleration
of the scale factor does not stop. These kind of potentials
are used as quintessence. Here, with the inclusion
of a negative term we could have an accelerating
scale factor but the period of positive acceleration
will necessarily stop.

 In fig.3 we
show the evolution for the same potential but
with $\dot\phi(t_0)=0$. We see that in this case
even though the potential is not symmetric
around $\phi(t_0)$ we have a symmetric solution
since   $\phi(t_0-t)=\phi(t_0+t)$.

\section{ Conclusions}

We have seen that a canonically normalized scalar field can lead
dynamically to a region with an oscillating equation of state
parameter with $\wp<-1$. For this to happen a
negative potential $V(\phi)$ is needed and the region
where $\wp<-1 \Leftrightarrow \rho_\phi <0$ and the WEC is
not violated.

During all the time that $\rho_\phi<0$ we have $|\rho_\phi|$,
smaller than the sum of all other energy densities
for a long period of time and regardless of the functional form
of the potential $V$. This opens the possibility of having
 a small cosmological constant today as the result of cancellation of higher
energy scales, as EW or susy breaking scales.

The total equation of state parameter $w$ increases its value with the
contribution of the scalar with $\wp<-1$ and $\rho_\phi<0$
since it gives a negative contribution to $\ddot a \sim -\rho_\phi(1+3\wp)$.

This work was supported in part by CONACYT project 32415-E and
DGAPA, UNAM project IN-110200.

\thebibliography{}

\footnotesize{

\bib{CMBR} {P. de Bernardis {\it et al}. Nature, (London) 404, (2000)
955, S. Hannany {\it et al}.,Astrophys.J.545 (2000) L1-L4}

\bib{SN1a} {A.G. Riess {\it et al.}, Astron. J. 116 (1998) 1009; S.
Perlmutter {\it et al}, ApJ 517 (1999) 565; P.M. Garnavich {\it et
al}, Ap.J 509 (1998) 74.}

\bib{neww} Carlo Baccigalupi, Amedeo Balbi, Sabino Matarrese,
Francesca Perrotta, Nicola Vittorio,  Phys.Rev. D65 (2002) 063520

\bib{cald}R.R. Caldwell, Phys.Lett. B545 (2002) 23.

\bib{mcin}B. McInnes,   JHEP 0208:029,2002 (astro-ph/0210321)
\bib{phantom} For an exhaustive list of reference see
P. Singh, M.Sami, N. Dadhich Phys.Rev. D68 (2003) 023522;  J. M.
Cline, S. Jeon, G. D. Moore, hep-ph/0311312.

\bib{ax.vuc} A. de la Macorra and  H. Vuccetich, in preparation

\bib{cyc} P.J. Steinhardt, N. Turok, Phys.Rev.D65:126003,2002

\bib{ekp} J. Khoury,   B. A. Ovrut, P. J. Steinhardt, N. Turok,
 Phys.Rev.D64:123522,2001

\bib{linde} A. Linde, JHEP 0111:052,2001; N. Felder, A.V. Frolov, L.
Kofman, A. V. Linde,  Phys.Rev.D66:023507,2002

\bib{chris1}{A. de la Macorra and C. Stephan-Otto,
 Phys.Rev.Lett.87:271301,2001
  Phys.Rev.D65:083520,2002 }

  \bib{tracker} I. Zlatev, L. Wang and P.J. Steinhardt, Phys. Rev.
Lett.82 (1999) 8960;  Phys. Rev. D59 (1999)123504

\bib{ax.mod} A. de la Macorra, JHEP 0301:033,2003  (hep-ph/0111292)

\bib{vilenkin} J. Garriga, A. Vilenkin,  Phys.Rev.D61:083502,2000

\bib{generic}  A.R. Liddle and R.J. Scherrer, Phys.Rev.
D59,  (1999)023509

\bib{mio.scalar}{A. de la Macorra and G. Piccinelli, Phys.
Rev.D61 (2000) 123503}

}
\end{document}